\documentclass[preprint,12pt,3p]{elsarticle}

\usepackage{matlab-prettifier}
\usepackage{graphicx}
\usepackage{makecell}
\usepackage{xfrac}
\usepackage{parskip}
\usepackage{adjustbox}
\usepackage{setspace}
\usepackage{lscape}
\usepackage[backref=page]{hyperref}
\hypersetup{ 
backref=true, 
bookmarks=true, 
pdftoolbar=true, 
pdfmenubar=true, 
pdffitwindow=false, 
pdfstartview={FitH}, 
pdftitle={}, 
pdfauthor={}, 
pdfsubject={}, 
pdfkeywords={}{}, 
pdfnewwindow=true, 
colorlinks=true, 
linkcolor=BlueViolet, 
citecolor=maroon, 
urlcolor=blue,
}
\usepackage{comment}
\usepackage{subfig}
\usepackage{caption}
\usepackage{longtable}
\usepackage{amssymb}
\usepackage{amsmath}
\usepackage{gensymb}
\usepackage{siunitx}

\usepackage{natbib}
\biboptions{comma,round}
\setcitestyle{authoryear}

\begin{document}
\begin{frontmatter}

\title{Database for the meta-analysis of the social cost of carbon (v2026.1)}

\author[label1,label2,label3,label4,label5,label6]{Richard S.J. Tol\corref{cor1}
}
\address[label1]{Department of Economics, University of Sussex, Falmer, United Kingdom}
\address[label2]{Institute for Environmental Studies, Vrije Universiteit, Amsterdam, The Netherlands}
\address[label3]{Department of Spatial Economics, Vrije Universiteit, Amsterdam, The Netherlands}
\address[label4]{Tinbergen Institute, Amsterdam, The Netherlands}
\address[label5]{CESifo, Munich, Germany}
\address[label6]{Payne Institute for Public Policy, Colorado School of Mines, Golden, CO, USA}

\cortext[cor1]{Jubilee Building, BN1 9SL, UK}

\ead{r.tol@sussex.ac.uk}
\ead[url]{http://www.ae-info.org/ae/Member/Tol\_Richard}

\begin{abstract}
A new version of the database for the meta-analysis of estimates of the social cost of carbon is presented. New records were added.
\end{abstract}

\end{frontmatter}

\section{Introduction}
This paper describes the databases underlying the meta-analysis of the social cost of carbon.

Previous meta-analyses of the social cost of carbon were published by \citet{Tol2005, Bergh2014, Havranek2015, Wang2019} and \citet{Moore2024PNAS}. My first attempt was updated in \citet{Tol2008Ejrn, Tol2009JEP, Tol2010PW, Tol2011, Tol2013JEDC, Tol2014bk, Tol2015macro, Tol2018REEP, Tol2019bk, Tol2020handbook, Tol2023bk, Tol2023NCC, Anthoff2022, Tol2025EE} and \citet{Tol2025anyas}. Many of the updates involve newer data only. The exceptions are \citet{Tol2008Ejrn} and \citet{Anthoff2022}, who focus on the tail of the distribution, \citet{Tol2018REEP} and \citet{Tol2025anyas} who try to detect publication bias (the latter successfully), and \citet{Tol2008Ejrn} and \citet{Tol2023NCC}, who study trends. \citet{TolTol2023} is about software for \href{https://richardtol.shinyapps.io/MetaSCC/}{online visualization} of meta-analysis.

The new version of the database includes new records.

Section \ref{sc:data} presents the databases. Section \ref{sc:results} shows descriptive statistics. Section \ref{sc:conclude} concludes. 

\section{Data}
\label{sc:data}

\subsection{Literature}
The current version of the database includes 528 papers, published in 156 journals, spanning the period from 1980 to 2025. These papers were collected over two decades. Only two papers are missing, the others are in the author's library, mostly in electronic form some only in hard copy. Papers were collected by systematic literature search, personal acquaintance, and exploration of the citations of previously considered papers.

This version of the database adds 60 papers published in 2025: \citet{Akdag2025, Aksoy2025, Arab2025, Ashenfarb2025, Behzadi2025, Bhattacharya2025, Bhave2024, Brooks2025, Caesary2025, Castellini2025, Chapin2025, Chen2025energy, Chen2025, Cheu2025, Cho2025, Coppens2025, Cunanan2025, daCosta2025, Deng2025, Eduardo2025, Estrada2025NComm, Estrada2025anyas, Fell2025, Feng2025, Folini2025, Garcia2025FR, Islam2025, Kashanian2025, Khan2025, Kerns2025, Kotchen2025, Laamouri2025, Landry2025, Lima2025, Lubis2025, Manan2025, Mohammadyari2025, Naef2025, Neal2025, Ovaere2025, Piazza2025, Ploeg2025, Pranita2025, Rao2025, Rose2025, Schaumann2025, Serfas2025, Springmann2025, Sreelekshmi2025, Wang2025, Wu2025FPE, Wu2025EEE, Wu2025CBM, Xiao2025, Xu2025, Xuan2025, Yu2025, Zhang2025, Zhu2025}.

I also added 21 previously overlooked papers: \citet{Al-Amin2015, Aleshina2024, Bilal2024, Canada2022, Deng2024, EDF2017, Eteriki2023, GAO2020, Howard2015a, Howard2015b, Landry2021, Li2025, Rasiah2016, Raynaud2014, Rose2022, Sarkar2019, Shelanski2015, Siagian2024, Sylvan2017, UBA2024, Wong2016}.

These paper add to 446 papers included in the previous version of the meta-analysis: \citet{Ackerman2006, AckermanMunitz2012, Ackerman2016, AckermanStanton2012, Ackerman2013, Adler2017, Agliardi2022, Allen2016, Anderson2014, Anthoff2007, Anthoff2009, Anthoff2019, Anthoff2009ere, Anthoff2010, Anthoff2011wp, Anthoff2013, Anthoff2014, Anthoff2022, Anthoff2009ee, Anthoff2009ejrn, Anthoff2009erl, Anthoff2011reg, Anthoff2011time, Asplund2017, Asplund2019, Audus1998, Ayres1991, Azar1994, Azar2023, Azar1996, Balmford2023, Barnett2020, Bastien-Olvera2021, Barrage2014, Barrage2018, Barrage2020, Barrage2020aer, Barrage2023, Barrage2024, Bauer2023, Belfiori2024, Below2014, Berger2017, Bertram2024, Bherwani2019, Bherwani2024, Bijgaart2013, Bijgaart2016, Bond2024, Botzen2012, Brander2010, Braun2024, Brausmann2024, Bremer2021, Bressler2021NComm, Bretschger2018, Bretschger2019, Brock2017, Brock2019, Budolfson2020, Budolfson2017, Budolfson2019, Caesary2023, Cai2019, Cai2012, Cai2015, Cai2016, Calvas2024, Ceronsky2005, Ceronsky2011, Chung2024, Clarkson2002, Cline1992, Cline1997, Cline2004, Coleman2021, Coppola2024, Crost2013, Crost2014, Czyz2023, Dangl2007, Daniel2019, Dayaratna2023, Dayaratna2017, Dayaratna2020, Dennig2014, Dennig2015, Dietz2011, Dietz2022, Dietz2021, Dietz2021JAERE, Dietz2015, Dietz2019, Dong2024, Downing1996, Downing2005, Drupp2021, Ekholm2018, EPA2023, EPA2009, Espagne2018, Eyre1999, Fankhauser1993, Fankhauser1994, Fankhauser1995, Farooq2024, Faulwasser2018, Faulwasser2018b, Fiestas-Chevez2024, Fillon2023, Foley2013, Freeman2015ej, Freeman2015jeem, Freeman2016, Fu2023, Gerlagh2023EER, Gerlagh2020, GerlaghLiski2017, GerlaghLiski2018, Gerlagh2023SJE, Gillingham2018, Glotter2014, Golosov2014, Golub2017, Gonzalez-Eguino2016, Gossling2024, Goulder2000, Greenstone2013, Griffiths2012, Groom2007, Groom2023, Groshans2019, Gschnaller2020, Guivarch2018, Guo2006, Hafeez2017, Haites1998, Hambel2021JIE, Hambel2021EER, Hambel2024, Hanley2004, Hansel2018, Hansel2020, Hansel2021, Haraden1992, Haraden1993, Hasan2024, Hassler2016, Hastunc2025, Hatase2015, Heal2014, HeinzowTol2003, Hepburn2009, HILLEBRAND2019, Hillebrand2023, Hoffman2024, Hohmeyer1996, Hohmeyer2002, Hohmeyer2004, Hohmeyer1992, Hong2023, Hope2005a, Hope2005b, Hope2005, Hope2006c, Hope2006a, Hope2006b, Hope2008a, Hope2008b, Hope2009, Hope2011, Hope2011wp, Hope2013, Hope2013book,  Hope2020, HopeHope2013, Hope1996, Howard2017, Howard2020, Howarth1998, Howarth2014, Hwang2013, Hwang2016, Hwang2017, Hwang2017CE, Hwang2019, IAWGSCC2010, IAWGSCC2013, IAWGSCC2015, IAWGSCC2021, Ibrahim2024, Ikefuji2021, Iverson2012, Iverson2015, Iverson2021, Jaakkola2022, Jaakkola2019,Hoffman2024,  Al-Jabir2024, Jensen2013, Jensen2014, Jensen2021, Jiang2024, Jin2024, Johnson2012, Kalkuhl2020, Karydas2019, Kaushal2023, Kellett2019, Khabarov2022, Kelly2001, Kemfert2010, Kessler2017, Kikstra2021, Knoke2023, Koch2023, Kon2024, Kopp2012, Kotchen2018, Kotlikoff2021, Krewitt2006, Kulkarni2024, Lemoine2015, Lemoine2021, Lemoine2014, Lemoine2016JEBO, Lemoine2016NCC, Li2016, Link2004, Lintunen2021scc, Liu2022, Lontzek2015, Loube2024, Lucchesi2024, Lupi2021, Maddison1994, Maddison1995, Manne2004lomb, Mardones2024, Marten2011, Marten2014, Marten2015, Marten2012, Marten2012wp, Marten2013, Mastandrea2001, Mendelsohn2003, Mendelsohn2004, Mendelsohn2005, Meng2024, Mensbrugghe2023, Mikhailova2019, Mikhailova2024a, Mikhailova2024, MnPUC2018, Molocchi2024, Moore2015, Moore2017, Moyer2014, Muangjai2024, Naeini2020, Narita2009, Narita2010jepm, Narita2010esp, Newbold2010, Newbold2013, Newbold2014, Newell2003, Newell2004, Newell2022, Niu2024, Nordhaus1980, Nordhaus1982, Nordhaus1989, Nordhaus1990, Nordhaus1990Brookings, Nordhaus1991MIT, Nordhaus1991AER, Nordhaus1991EJ, Nordhaus1992, Nordhaus1993AER, Nordhaus1993JEP, Nordhaus1993, Nordhaus1994book, Nordhaus1994IIASA, Nordhaus1997IPIECA, Nordhaus2001, Nordhaus2007jel, Nordhaus2007reep, Nordhaus2007sci, Nordhaus2008, Nordhaus2009, Nordhaus2010, Nordhaus2011, Nordhaus2013, Nordhaus2013iam, Nordhaus2014, Nordhaus2015, Nordhaus2017, Nordhaus2017rep, Nordhaus2018aej, Nordhaus2018cc, Nordhaus2019pnas, Nordhaus2019aer, NordhausBoyer2000, NordhausPopp1997, NordhausSztorc2013, NordhausYang1996, Nyserda2021, Okullo2020, Olijslagers2022, Olijslagers2023, Olijslagers2024, Ortiz2011, Otto2013, Parry1993, Pearce2003, Pearce2005, Peck1992, Peck1993, Peck1994, Peck1995, Peck1996RA, Penner1992, Perrissin2012, Pezzey2014, Pindyck2017coase, Pindyck2019, Piontek2019, Pizer1999, Pizer2002, PlambeckHope1996, Ploeg2014EER, Ploeg2015, Ploeg2016wp, PloegRezai2017, PloegRezai2018, PloegRezai2019eer, PloegRezai2019, Rezai2012wp, Ploeg2018, Ploeg2021, Ploeg2021ITPF, PloegZeeuw2013, PloegZeeuw2014, PloegZeeuw2016, PloegZeeuw2018, Ploeg2019, Poelhekke2019, Popp2004, Portland2021, Pottier2015, Prest2023, Prest2024, Price2007, Pulhin2024, Pycroft2011, Pycroft2014, Quiggin2018, RAUTIAINEN2017, Reilly1993, Rennert2019, Rennert2021, Rennert2022, Rezai2010, Rezai2015, RezaiPloeg2016, RezaiPloeg2017, RezaiPloeg2017ere, RezaiPloeg2017ms, Rezai2012, Rezai2020, Ricke2018, Rickels2023, Rickels2024, Rose2017, Roughgarden1999, Rudik2020, Russell2022, DiRusso2025, Safarzynska2022, Schauer1995, Schmidt2013, Schultes2021, Scovronick2017, Scovronick2020, Shafiei-Alavijeh2024, Sen2024, Shindell2015, Shindell2017, Smith2023, Sohn2019, Sohngen2010, Stern2006, Stern2007sci, Striepe2024, Su2017, Su2024, Taconet2021, Tao2024, Tian2019, Tibebu2021, Tibebu2024, Tol1999marg, Tol2003, Tol2005EDE, Tol2005OUP, Tol2010lomborg, Tol2010, Tol2012ERE, Tol2013el, Tol2013wp, Tol2019EE, Tol2000, Tol2001, Tol2020handbook, Tol2024EnPol, Traeger2014, Traeger2015, Traeger2023, UBA2007, UBA2012, UBA2019, UBA2020, Uzawa2003, Wahba2006, Waldhoff2011, Waldhoff2014, Wang2022OE, Wang2024DA, Watkiss2005, Watkiss2008, Weitzman2013, Weller2015, Wijst2021, Wijst2023, Wong2015, Wu2024, Yang2018, Yilma2024, Yohe2013, Yoo2018, Yoo2021, Zhao2023, Zhen2020} and \citet{Zhen2018}.

\subsection{Main database}
Results and assumptions were coded for each of the 528 papers. In most cases, tables and text were used. Some papers present results only graphically, in which case authors were contacted by email. If there was no response\textemdash in one case, the authors responded that their hard disk had crashed\textemdash graphical results were digitized with the \textsc{grabit} function in \textsc{Matlab}.

The main database is in \textsc{Excel}. It consists of the following fields, for each of the 14,884 estimates of the social cost of carbon:
\begin{itemize}
    \item Year of publication.
    \item Author weight, indicating the apparent weight the authors give to each estimate.
    \item Paper weight, equal to one over the number of estimates contained in the paper.
    \item Quality weight, equal to one plus:
    \begin{itemize}
        \item Peer-review, a dummy variable indicating whether the paper was independently reviewed;
        \item Method, a dummy variable indicating whether the social cost of carbon was calculated or approximated using a mathematically accepted method; and
        \item Scenario, a dummy variable indicating whether not-implausible scenarios were used.
    \end{itemize}
    \item Censor, a variable equal to one if the estimate is lower than \$1,594/tC, the Leviathan tax in 2019 \citep{Tol2012CCL}, and zero if it is greater than \$11,571/tC, the maximum ability to pay. The variable is scaled linearly between 0 and 1 for values in between.
    \item Social cost of carbon, as reported.
    \item Emission year.
    \item Dollar year.
    \item Unit, indicating whether the social cost of carbon is expressed in metric tonnes\footnote{No study reports non-metric units.} of carbon or of carbon dioxide.
    \item \textcolor{red}{Legacy} The social cost of carbon for 2010 in 2010 US dollar per metric tonne of carbon.
    \item The social cost of carbon for 2025 in 2024 US dollar per metric tonne of carbon.
    \item Negative, a dummy variable whether or not the study permits the possibility of a social benefit of carbon.
    \item Consumption discount rate.
    \item Pure rate of time preference.
    \item Rate of risk aversion.
    \item Inverse of the elasticity of intertemporal substitution.
    \item Rate in inequity aversion.
    \item Assumed \emph{level} impact of 2.5\celsius{} warming.
    \item Functional form of the level damage function.
    \item Assumed \emph{growth} impact of 1.0\celsius{} warming.
    \item Functional form of the growth damage function.
    \item Global warming in 2100.
    \item Equity weights, a dummy variable whether or not used.
    \item Mean, a dummy variable whether the expectation of the social cost of carbon is reported (1) or some other statistic (0).
    \item Stochastic, a dummy variable whether the model used is stochastic (1) or not.
    \item Analytic, a dummy variable whether the model is solved analytically (1) or numerically (0).
    \item Dynamic, a variable indicating whether vulnerability to climate change falls with development (1), increases (-1), or does not change (0).
    \item Pigou, a dummy variable whether the social cost of carbon is imposed on emissions or not.
    \item Paper ID.
\end{itemize}

The data are on \href{https://github.com/rtol/metascc/tree/master}{GitHub}, file socialcostcarbon.xlsx.

A supplementary database records the social cost of carbon for different years of emission. This is used to calculate the \emph{population} average growth rate, which is 2.07\% per year in the current database. This growth rate is used to shift the social cost of carbon for the year of reporting to 2025.

A \emph{study-specific} growth rate is computed too. This is used, where available, to compute an alternative social cost of carbon for 2025.

These data are in the same file as above.

\subsection{Satellite data}
There are four satellite databases:
\begin{enumerate}
    \item Bibliography, in \textsc{BibTex}. This preserves the sanity of the author and allows others to replicate the research. This database contains:
    \begin{itemize}
        \item Paper ID.
        \item Title.
        \item Authors.
        \item Publication year.
        \item Journal name / volume title / publication series.
    \end{itemize}
    The data are on  \href{https://github.com/rtol/metascc/tree/master}{GitHub}, file scc.bib
    \item Co-author network, in \textsc{Matlab} and \textsc{Excel}. This allows for testing idiosyncratic preferences and influence. This database contains:
    \begin{itemize}
        \item Paper ID.
        \item Author ID.
        \item Gender (at the time of publication).
        \item A dummy variable indicating whether a particular researcher authored a particular paper.
    \end{itemize}
    The data are on \href{https://github.com/rtol/metascc/tree/master}{GitHub}, files citation.m and connect.xlsx.
    \item Country of affiliation, in \textsc{Excel}. This allows for testing representativeness. This database contains:
    \begin{itemize}
        \item Paper ID.
        \item Country ID.
        \item A dummy variable indicating whether a particular paper was authored by a researcher based in a particular country.
    \end{itemize}
    The data are on \href{https://github.com/rtol/metascc/tree/master}{GitHub}, file connect.xlsx
    \item Citation network, in \textsc{Matlab}. This allows for testing influence and citation bias. This database contains:
    \begin{itemize}
        \item Paper ID.
        \item Citation ID.
        \item Citation ID dyads.
    \end{itemize}
    The data are on \href{https://github.com/rtol/metascc/tree/master}{GitHub}, file citation.m
\end{enumerate}

The five databases have the field \textit{Paper ID} in common and can therefore readily be combined in any constellation.

\section{Descriptive statistics}
\label{sc:results}
\begin{figure}
    \centering
    \includegraphics[width = \textwidth]{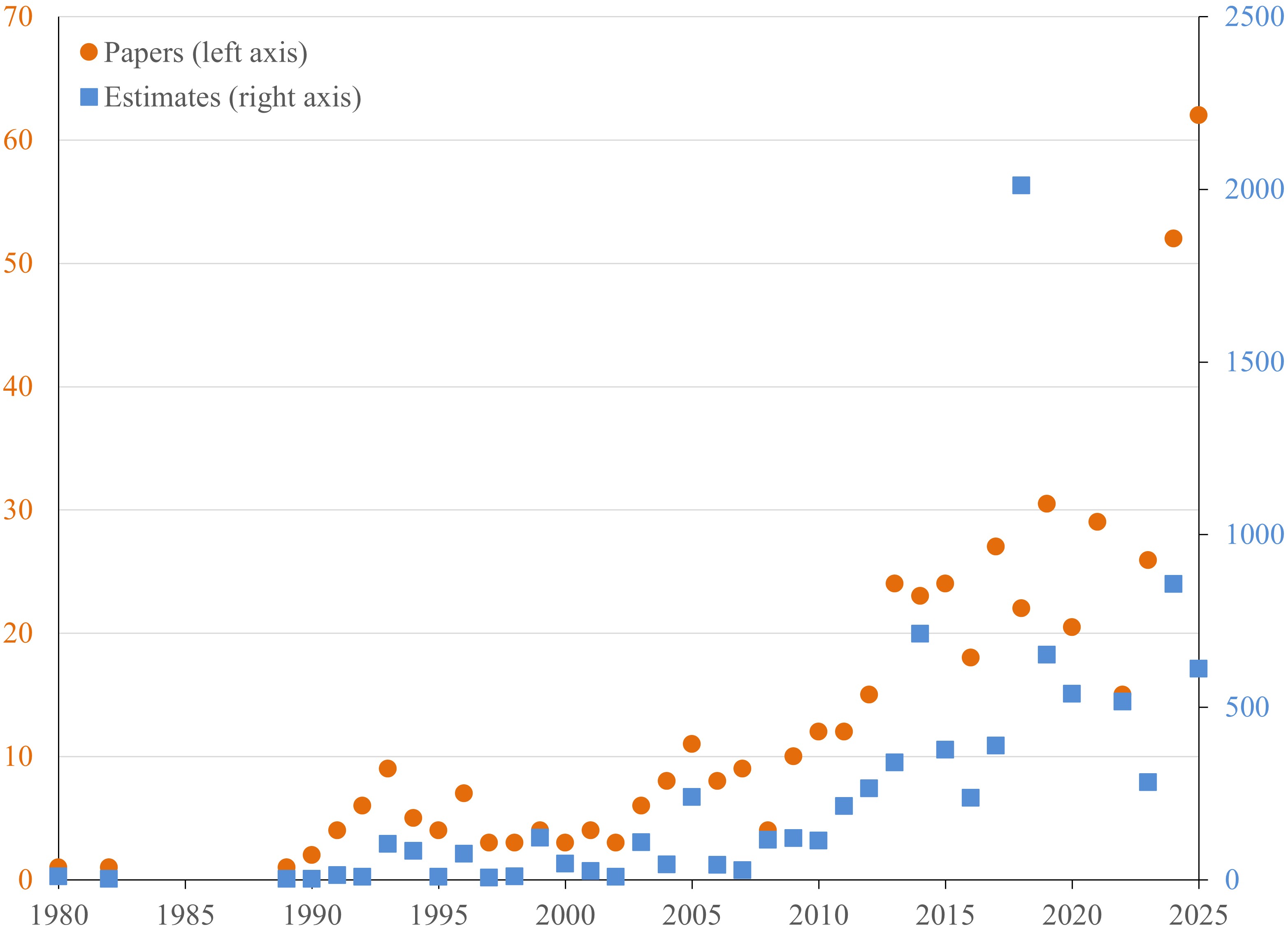}
    \caption{The number of papers on and estimates of the social cost of carbon by year.}
    \label{fig:number}
\end{figure}

Figure \ref{fig:number} shows the number of papers and the number of estimates per publication year. The first two papers were published in 1980 and 1982, with 12 estimates between them. More papers were published in the early 1990s. Since 2010, ten papers or more were published per year. The record was set in 2025 with 62 new papers. The number of estimates increased more rapidly than the number of papers as computers got faster, referees more demanding, and supplementary information more extensive. The maximum was reached in 2021, with 5,511 estimates reported (in 27 papers).\footnote{There are many more unpublished results. Chris W. Hope routinely runs Monte Carlo analysis with the \textsc{page} model but never reports the full results. \citet{Rudik2020} is based on 50,000 Monte Carlo runs, \citet{Anthoff2022} on 100,000 each for three alternative models.} 

\begin{figure}
    \centering
    \includegraphics[width = \textwidth]{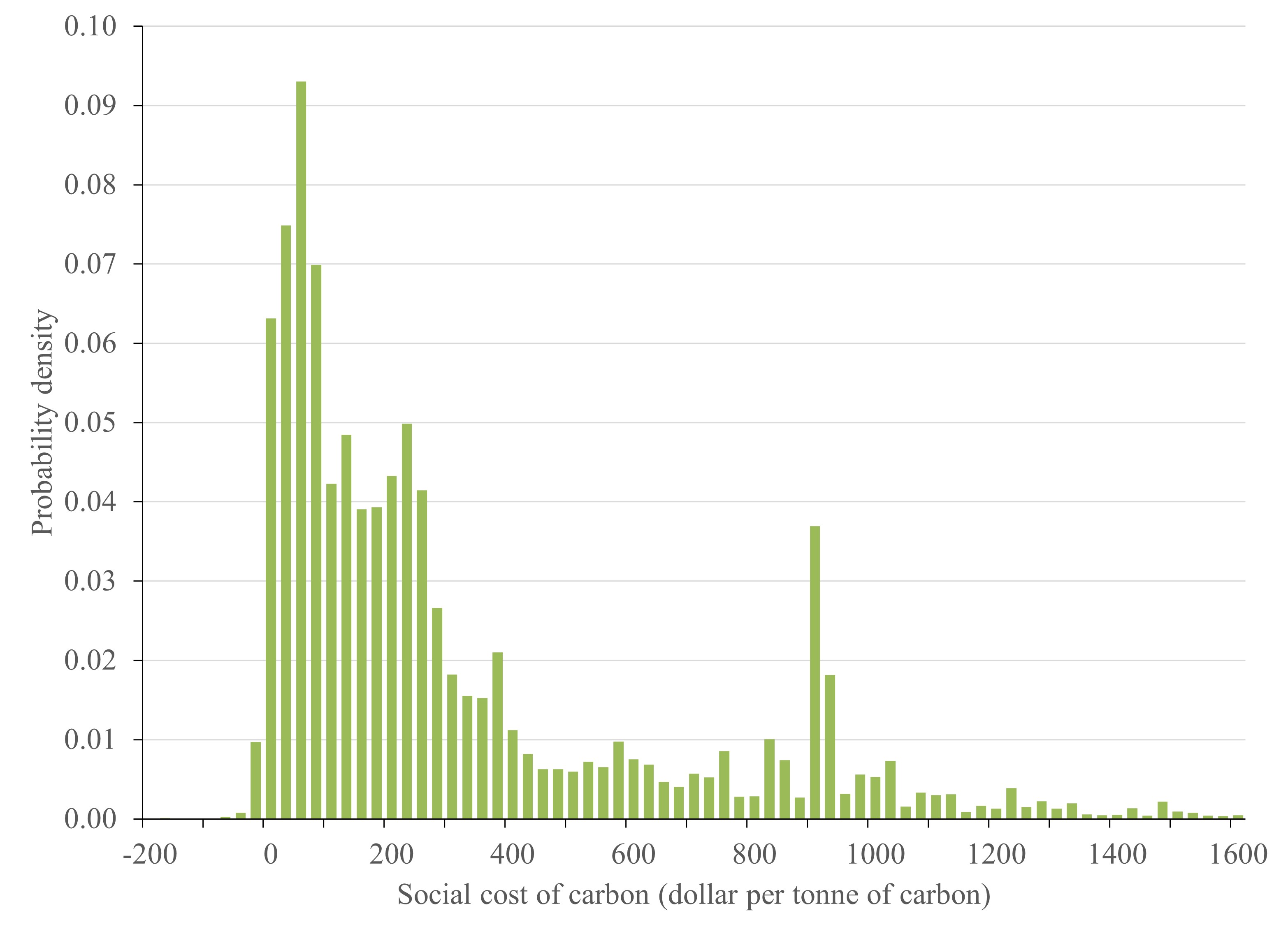}
    \caption{The histogram of published estimates of the social cost of carbon. Estimates are author- and quality-weighted and censored.}
    \label{fig:histo}
\end{figure}

Figure \ref{fig:histo} shows the histogram of all estimates of the social cost of carbon. The mode of the empirical distribution lies between \$50/tC and \$75/tC. A second mode lies between \$900/tC and \$925/tC, the oft-repeated estimate of \citet{Rennert2022}. The distribution is skewed, with a pronounced right tail. The mean is thus much larger than the mode: \$411/tC (s.e. \$10/tC). The number of estimates showing a social \emph{benefit} of carbon is small, only 1.1\% of the total.

\begin{figure}
    \centering
    \includegraphics[width = 0.49\textwidth]{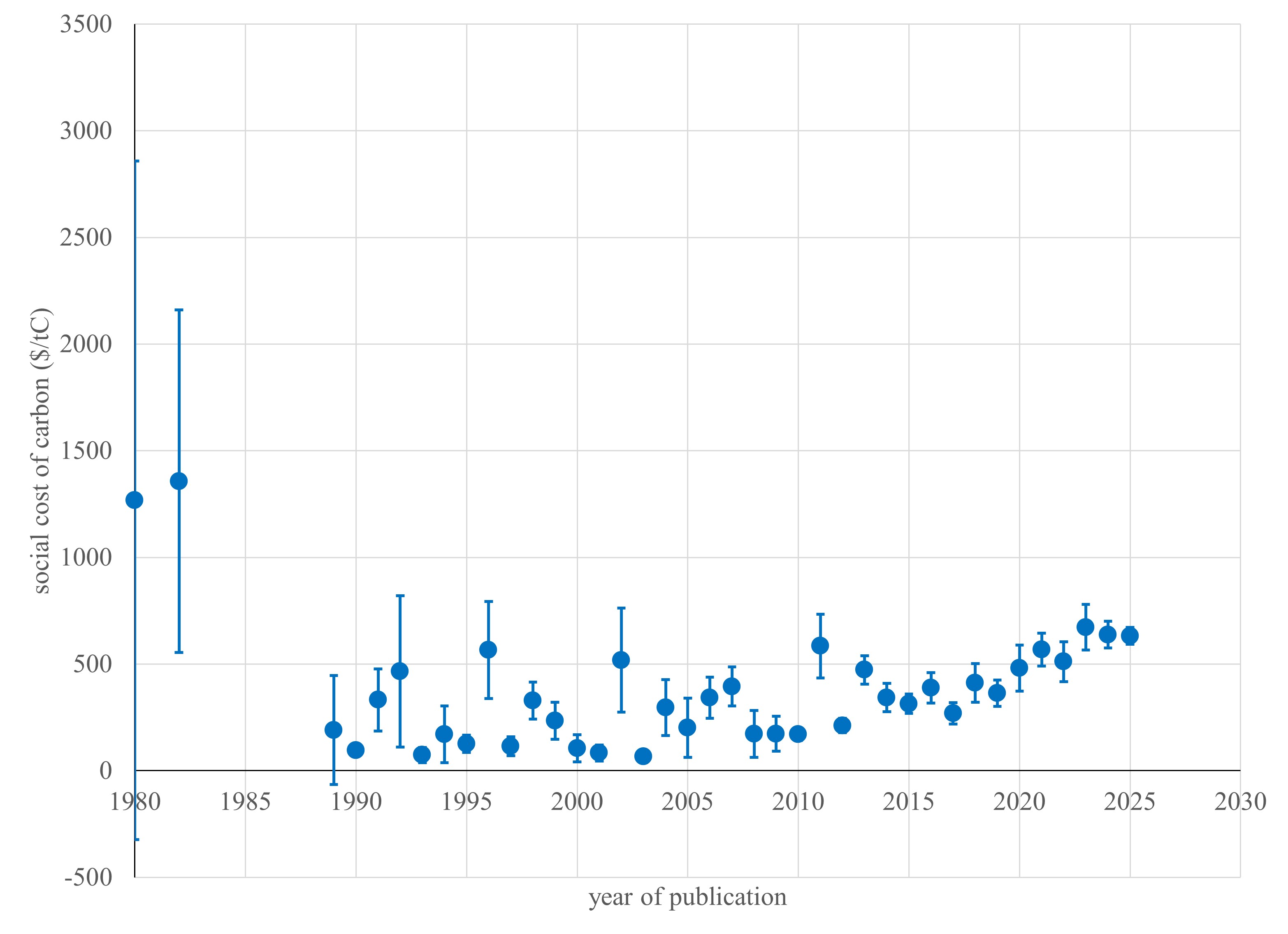}
    \includegraphics[width = 0.49\textwidth]{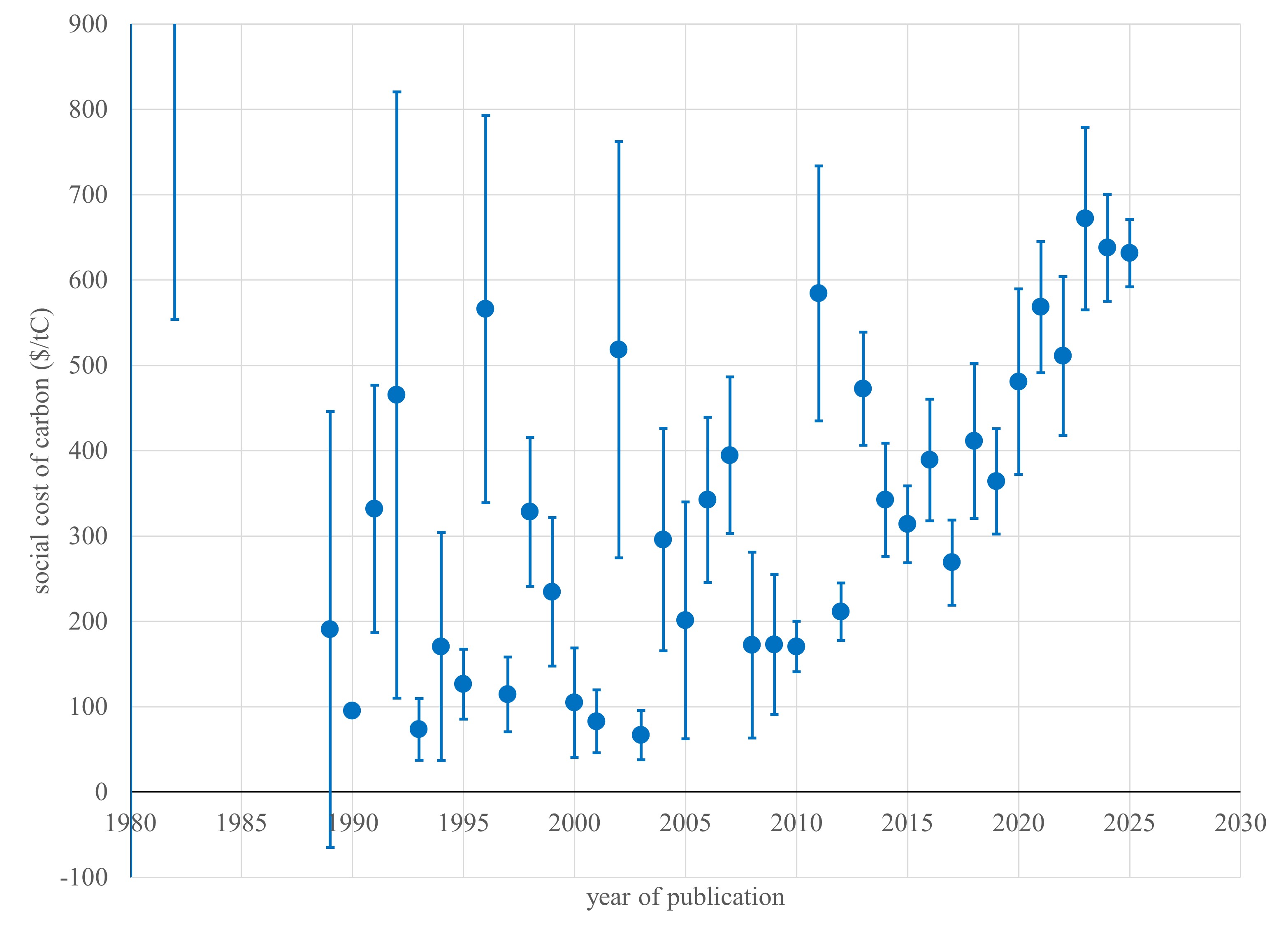}
    \caption{The average social cost of carbon by year of publication. The interval shown is the mean plus and minus the standard deviation. Estimates are author- and quality-weighted and censored. Both panels show the same data, but the right panel uses an abridged vertical axis.}
    \label{fig:time}
\end{figure}

Figure \ref{fig:time} shows the mean and range of estimates of the social cost of carbon over time. Early estimates were very high. Later estimates were lower but quite variable from year to year. There appears to be an upward trend from 2010 to 2023.

A refined field reveals that 81.5\% of estimates of the social cost of carbon assume that vulnerability to climate change is independent of development \citep[cf.][]{Schelling1984}. 17\% (1.5\%) assume that vulnerability falls (rises) with income. The social cost of carbon is \$124/tC lower (\$47/tC higher) than with constant vulnerability.

\begin{figure}
    \centering
    \includegraphics[width = \textwidth]{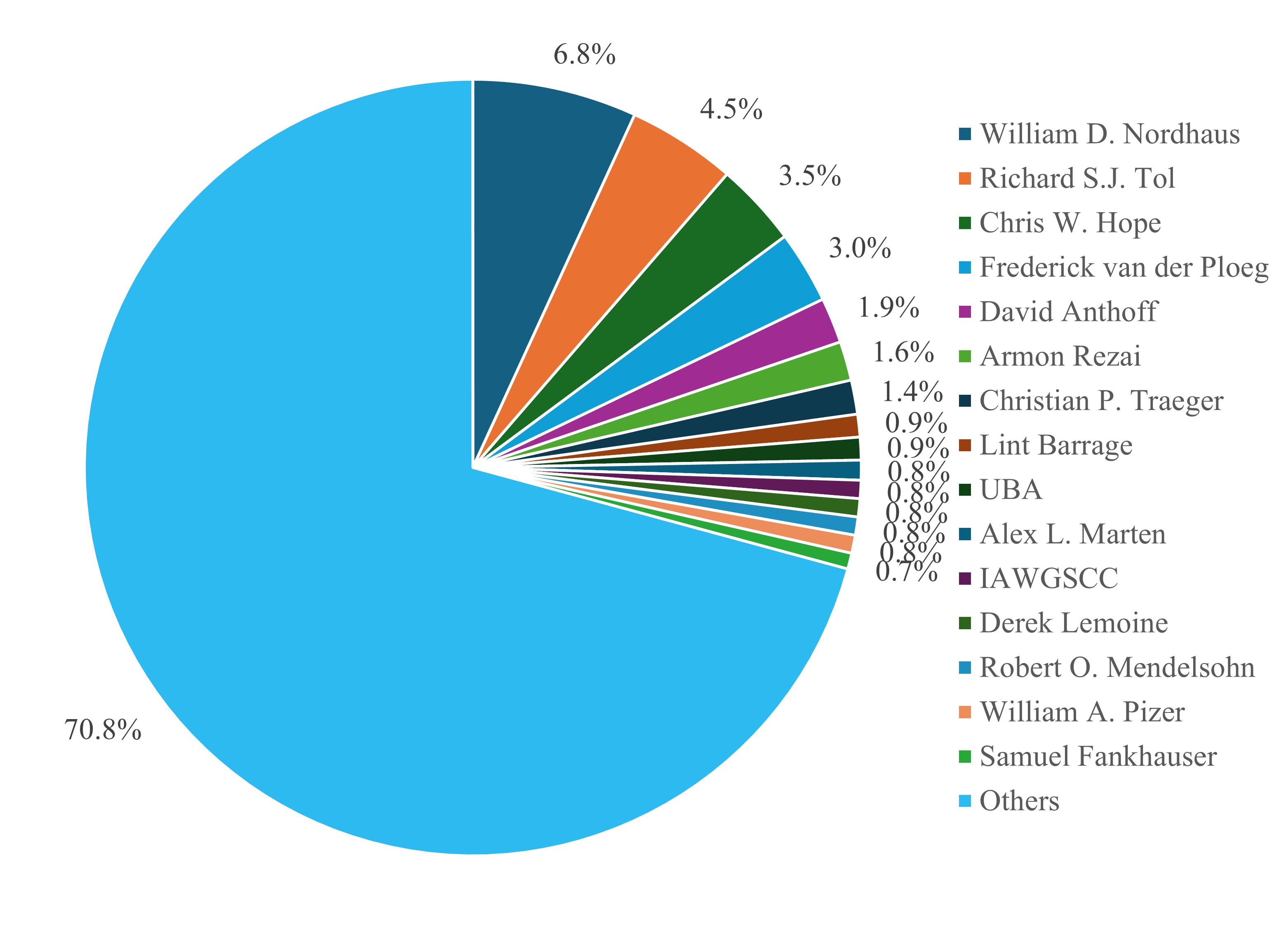}
    \caption{Contribution to the literature on the social cost of carbon by author.}
    \label{fig:author}
\end{figure}

Eight hundred seventy-four authors contributed to the literature on the social cost of carbon. The number of authors per paper varies between 1 and 29 \citep{Bertram2024}. The average paper has 1.66 authors. Figure \ref{fig:author} shows the 15 authors who contributed most to the literature on the social cost of carbon. Papers are shared equally between authors. That is, if a paper has two (three) authors, both (each) get assigned a half (third) paper, so that the sum of author-contributions equals the number of papers. Richard S.J. Tol (co-)authored most papers (46) but William D. Nordhaus, who tends to work alone or with one junior, has made the largest contribution (6.8\%). The ten most prolific authors together cover only 25.5\% of the literature.

The average estimate for female authors is \$51/tC higher than for males. The average estimate in economics journals is \$137/tC lower than in other journals.

\begin{figure}
    \centering
    \includegraphics[width = \textwidth]{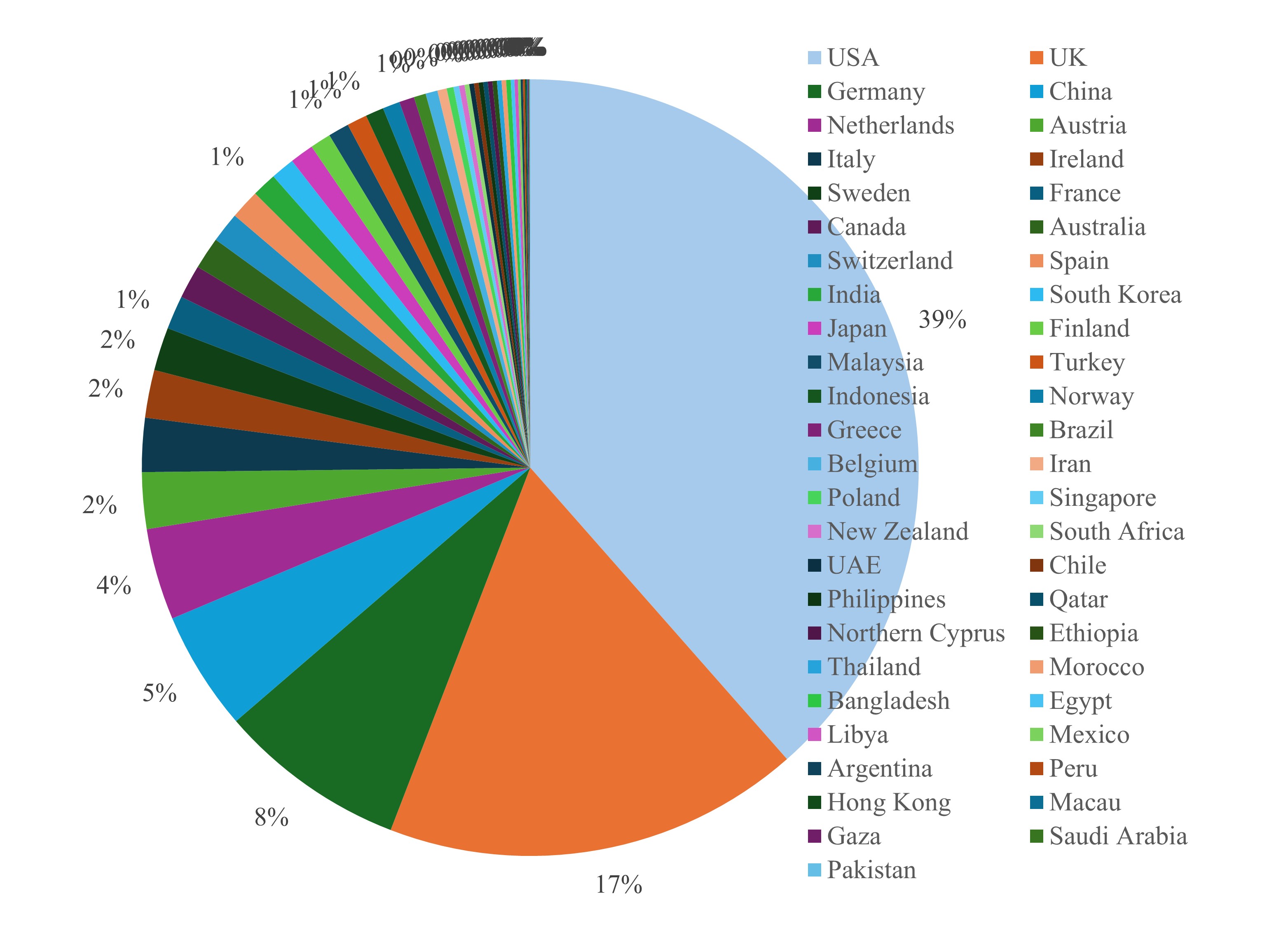}
    \caption{The number of papers on the social cost of carbon by country of affiliation.}
    \label{fig:country}
\end{figure}

Figure \ref{fig:country} shows the same information but for the country of affiliation. Some 38.5\% of all papers on the social cost of carbon were authored by researchers based in the USA, 17.3\% in the UK, 7.8\% in Germany, and 5.0\% in China. Forty-five other countries contributed to this literature. The remaining 140-odd countries did not. There are few papers from authors in Latin America and Africa.

Figure \ref{fig:coauthor} shows the co-author networks of everyone who has published five papers or more on the social cost of carbon. Authors of six papers or more are named. These seven networks together include N out of 872 authors. The smallest network is Stephen C. Peck and Thomas J. Teisberg. The largest network has N members, including most of the most prolific authors (cf. Figure \ref{fig:author}).

Figure \ref{fig:citation} shows the citation network. The node size is weighted arithmetic incloseness. Incloseness is a measure of influence, counting not just the citations of a paper, but also citations of citations, citations of citations of citations, and so on. Incloseness is an average distance. In this case, the arithmetic average is used, or rather a weighted average where the weights equal one over the number of citations. That is, if a paper is the only citation in another paper, the weight is one. If a paper is one of two citations, the weight is one-half. And so on. Figure \ref{fig:citation} reveals a dense and complicated citation network. Because papers are often circulated as working papers prior to publication, there are cycles, short ones\textemdash a paper citing a paper that cites it\textemdash as well as long ones\textemdash up to nine papers. The most influential papers are named. These were identified by running a regression of incloseness on the year of publication. The 5\% papers that are furthest from the regression line are named.

\section{Conclusion}
\label{sc:conclude}
This paper presents the 2026 version of the database of estimates of the social cost of carbon. Some records were updated and 81 new records added. The 2026 database is considerably larger than the 2025 one: 528 vs 446 papers. The average social cost of carbon of previously included studies is \$400/tC; the new studies average \$629/tC.

The database will be updated with newer publications on the social cost of carbon and inadvertently overlooked older publications. New fields may be added as necessary.

\newpage \bibliography{master}

\begin{landscape}
\begin{figure}
    \centering
    \includegraphics[width = 1.2\textwidth]{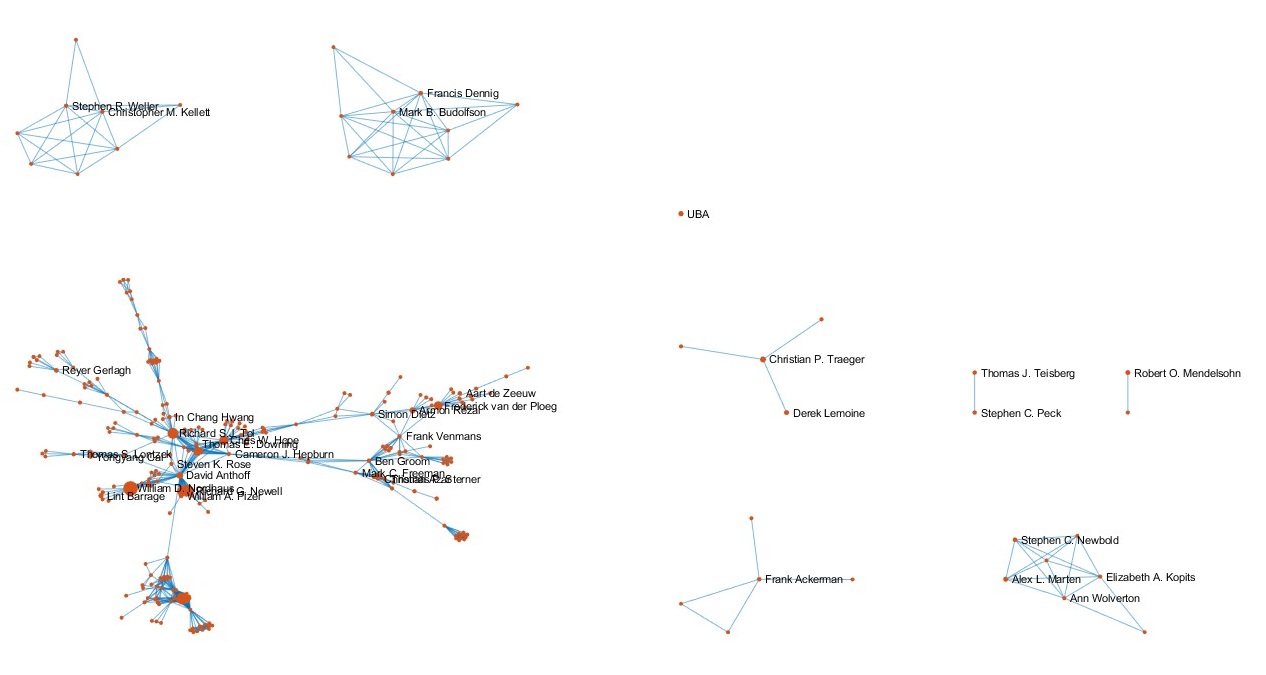}
    \caption{The nine most prolific co-author networks. Node size is the contribution to the literature. Authors of five papers or more are named.}
    \label{fig:coauthors}
\end{figure}

\begin{figure}
    \centering
    \includegraphics[width = 1.2\textwidth]{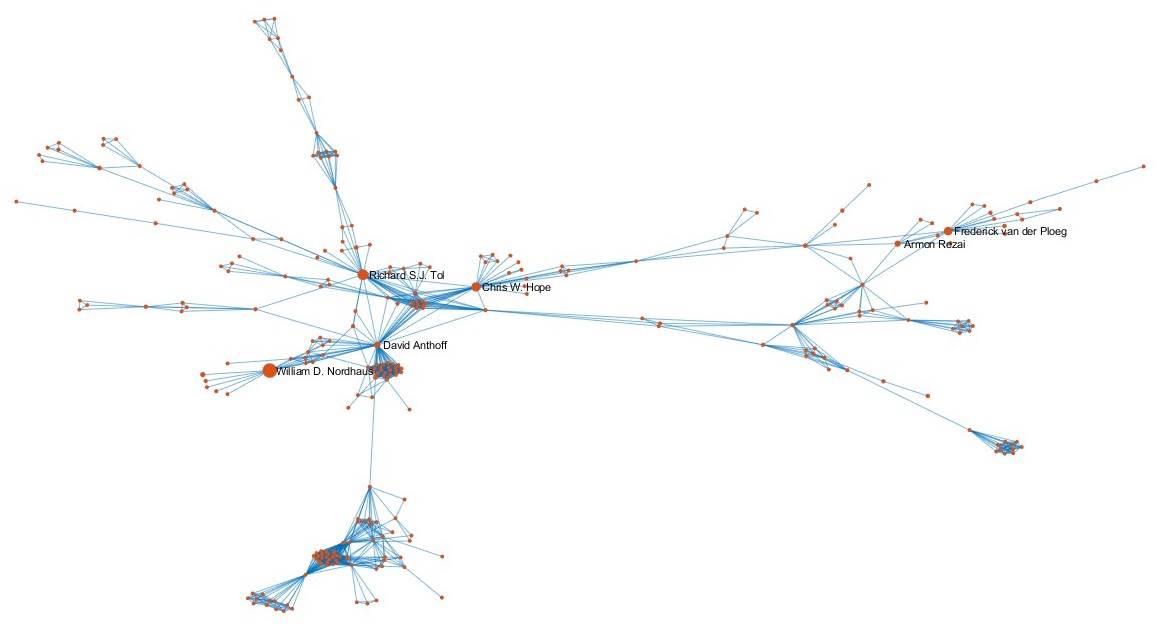}
    \caption{The most prolific co-author network. Node size is the contribution to the literature. Authors of thirteen papers or more are named.}
    \label{fig:coauthor}
\end{figure}

\begin{figure}
    \centering
    \includegraphics[width = 1.2\textwidth]{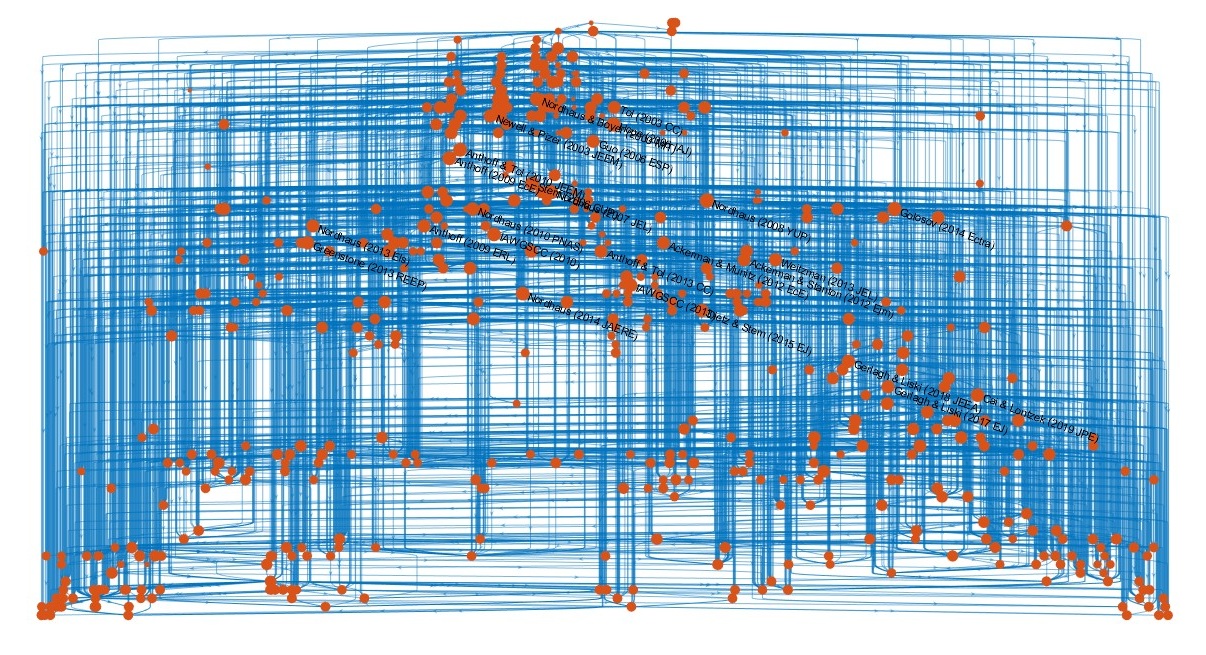}
    \caption{The citation network of papers on the social cost of carbon. Node size is incloseness corrected for number of citations and age. Top 5\% papers are named.}
    \label{fig:citation}
\end{figure}
\end{landscape}

\end{document}